\documentclass[preprint,12pt]{elsarticle}

\usepackage{lineno,hyperref}
\usepackage{color,ulem}
\modulolinenumbers[1]

\usepackage{amsmath} 

\journal{Astroparticle Physics}

\begin{document}

\begin{frontmatter}

\title{Evaluations of uncertainties in simulations of propagation of ultrahigh-energy cosmic-ray nuclei derived from microscopic nuclear models}





\author[1]{E.~Kido} 
\author[2]{T.~Inakura}
\author[3]{M.~Kimura}
\author[4]{N.~Kobayashi}
\author[1,5,6]{S.~Nagataki}
\author[7]{N.~Shimizu}
\author[4]{A.~Tamii}
\author[8]{Y.~Utsuno}

\address[1]{RIKEN Cluster for Pioneering Research,\\
Astrophysical Big Bang Laboratory (ABBL),
  Wako, Saitama, Japan}

\address[2]{Laboratory for Zero-Carbon Energy, Tokyo Institute of Technology,\\
Meguro, Tokyo, Japan}

\address[3]{Hokkaido University,\\
Sapporo, Hokkaido, Japan}

\address[4]{Research Center for Nuclear Physics, Osaka University,\\
Ibaraki, Osaka, Japan}

\address[5]{RIKEN Interdisciplinary Theoretical and Mathematical Sciences Program (iTHEMS),\\
Wako, Saitama, Japan}

\address[6]{Astrophysical Big Bang Group (ABBG),\\
Okinawa Institute of Science and Technology Graduate University (OIST),\\ 
Tancha, Onna-son, Kunigami-gun, Okinawa, Japan}

\address[7]{Center for Computational Sciences, University of Tsukuba,\\
 Tsukuba, Ibaraki, Japan}

\address[8]{Japan Atomic Energy Agency,\\
Tokai, Ibaraki, Japan}


\begin{abstract}
Photodisintegration is a main energy loss process for ultrahigh-energy cosmic-ray (UHECR) nuclei in intergalactic space. Therefore, it is crucial to understand systematic uncertainty in photodisintegration when simulating the propagation of UHECR nuclei. In this work, we calculated the cross sections using the random phase approximation (RPA) of density functional theory (DFT), a microscopic nuclear model. 
We calculated the $E1$ strength of 29 nuclei using three different density functionals.
We obtained the cross sections of photonuclear reactions, including photodisintegration, with the $E1$ strength. Then, we implemented the cross sections in the cosmic-ray propagation code CRPropa. We found that assuming certain astrophysical parameter values, the difference between UHECR energy spectrum predictions using the RPA calculation and the default photodisintegration model in CRPropa can be more than the statistical uncertainty of the spectrum. We also found that the differences between the RPA calculations and CRPropa default in certain astrophysical parameters obtained by a combined fit of UHECR energy spectrum and composition data assuming a phenomenological model of UHECR sources can be more than the uncertainty of the data.
\end{abstract}


\end{frontmatter}


\section{Introduction}

Observations of ultrahigh-energy cosmic rays (UHECRs) recently showed indications that some extragalactic sources emit atomic nuclei heavier than protons at the highest energies. 
The energy dependence of the mean values of the slant depth at the shower maximum $\langle X_{\rm max} \rangle$, which the Pierre Auger Observatory~\cite{AugerNIM} detected, was not compatible with the expectation of pure protons or pure iron when three hadron interaction models tuned to the LHC data are used~\cite{AugerComp}.
The result implied that the composition of UHECRs becomes lighter up to $10^{18.27}$ eV and heavier above that energy.
The contribution of light nuclei between protons and iron was needed to interpret the observed distribution of the $X_{\rm max}$~\cite{AugerComp2}. 
A significant dipole amplitude of a large-scale anisotropy in the arrival directions of UHECRs was discovered by the Pierre Auger Collaboration, and the direction of the dipole was compatible with the source distribution in the local large-scale structure~\cite{AugerDipole, dipoleupdate1, dipoleupdate2}.
Some evidence of the correlation between the arrival directions of UHECRs and extragalactic sources was detected by the Pierre Auger Observatory~\cite{AugerSBG} and by a working group~\cite{AugerTASBG} of Auger and Telescope Array~\cite{TANIM}.

The experimental evidence induces motivation to study light nuclei emitted from extragalactic sources.
To constrain astrophysical parameters such as the power index of the injection spectrum and acceleration limit of possible extragalactic sources, fitting observables such as energy spectrum and $X_{\rm max}$ using simulations of the propagation of UHECR nuclei in intergalactic space is useful.
The most critical interactions of the nuclei at the highest energies are photonuclear reactions with cosmic microwave background (CMB) photons.
The average energy of a CMB photon is approximately $6\times10^{-4}$~eV today. The energies of CMB photons reach a few tens of MeV in the rest frame of UHECR nuclei with Lorentz factors greater than approximately $10^{9}$. If a photon with a few tens of MeV in the rest frame of a nucleus is absorbed by the nucleus, the nucleus is excited by the photon. Accelerator experiments measured the photonuclear reactions in the laboratories. The electric dipole ($E1$) excitation is the dominant component of the excitation.
There are large peaks in the cross sections at approximately 20 MeV known as giant dipole resonances (GDRs), which are dominant in the $E1$ excitation.
The excited nucleus loses energy by emitting $p$, $n$, $\alpha$, $\gamma$ particles and so on.
The impact of the GDRs on UHECR nuclei can be found for example in Ref.~\cite{Allard}, Fig.~3 (right). 

However, the cross sections of photonuclear reactions often suffer from systematic discrepancies among different experimental methods at different accelerator facilities. In one extreme case~\cite{IAEA}, the systematic differences in the neutron-emitting cross sections of 19 nuclei between the measurements in Livermore and Saclay were up to a factor of two.
There is also a lack of measurements for many elements~\cite{DESYConcept}.
Furthermore, the cross sections of light nuclei are difficult for nuclear theories to describe accurately because individual nuclei manifest their own complicated aspects of nuclear structure, e.g., shell structure, deformation, and $\alpha$ cluster structure. 
Solving these issues in both experiments and theories 
is one of the main motivations of the Photo-Absorption of Nuclei and Decay Observation for Reactions in Astrophysics (PANDORA) project~\cite{PANDORA, PANDORApaper}.
Three facilities in Japan, Romania, and South Africa plan to systematically measure photoabsorption cross sections and $p$, $n$, $\alpha$, and $\gamma$ emission from light to $A \sim 60$ nuclei in this project with modern experimental methods of virtual-photon excitation by proton scattering~\cite{von2019electric} and real-photon excitation by a high-intensity laser-Compton scattering gamma-ray beam~\cite{gales2018extreme}.
This project will test the consistency among three facilities ensuring the mutual consistency through measurements on $^{27}$Al targets.

The Puget--Stecker--Bredekamp (PSB) model~\cite{PSB}, which implements a single decay chain for each nucleus, was widely used.
In Ref.~\cite{Khan}, it was recommended for simulations of UHECRs around 10$^{21}$ eV to implement multiple decay chains using TALYS~\cite{TALYS}, which is now used in several propagation codes of UHECRs, e.g., CRPropa~\cite{CRPropa}, SimProp~\cite{SimPropv2r4}, and \textsc{PriNCe}~\cite{DESY}.
The impact of uncertainties of photodisintegration cross sections on the simulations of UHECRs was evaluated in Ref.~\cite{DESY,SimProp, AugerFit}. The model dependence of the PSB model, PEANUT~\cite{PEANUT}, and TALYS was evaluated in Ref.~\cite{DESY}. The PSB model and parametrization of Kossov~\cite{Kossov} were compared with TALYS, and the impact of the difference of $\alpha$-particle emitting cross sections was also evaluated in Ref.~\cite{SimProp}. In Ref.~\cite{AugerFit}, the difference between TALYS, PSB, and Geant4~\cite{Geant4} models was shown.
TALYS can include $E1$ strength functions which are directly converted to the cross sections used in these propagation codes. 
If no data on $E1$ strength is included in TALYS, these cross sections are calculated using the phenomenological empirical laws; the peak energy of the cross sections is given as $(31.2 A^{-1/3}+20.6 A^{-1/6})$ MeV. 
It is known that this empirical law overestimates the peak energy in light nuclei, while it provides reasonable values for heavy nuclei.
For some nuclei (nine nuclei in light nuclei to $A\sim 60$), peak energies that are evaluated from the experimental data are used for the calculations. Otherwise, the empirical laws are used. Also nuclear deformation affects the cross section. The deformation splits the cross section distribution to a two-peak structure and shifts the peak energy by 1-2 MeV from one in the spherical case. This effect is not taken into account in the empirical law. The cross sections calculated by TALYS have non-negligible uncertainties. 
Therefore, the propagation codes mainly use the current experimental data for $E1$ strength functions of TALYS\@.

The PANDORA project plans to measure a few nuclei in a few years and measure a few dozens of important nuclei in the next decade for systematically evaluating and improving the model predictions~\cite{PANDORApaper}. 
The photonuclear reactions will be predicted by the models for the rest of the nuclei relevant to the UHECR propagation.
Therefore, the reliable nuclear model prediction is indispensable for the project and would also be informative to interpret the current experimental data.
In this work, we calculated the cross sections within the random phase approximation (RPA) of nuclear-density functional theory (DFT)~\cite{Inakura1, Inakura2},
which describes the $E1$ mode as a harmonic vibration mode of the nuclear potential.
The RPA is the most standard approach to microscopically calculate the $E1$ strength in nuclear theories.
Other nuclear theories than DFT such as {\it ab initio} type calculations, large-scale shell-model calculations~\cite{Shellmodel}, and Antisymmetrized Molecular Dynamics (AMD)~\cite{AMD1, AMD2} were reviewed in Section~3 in Ref.~\cite{PANDORApaper}.
In particular, the RPA calculations can be applied to a wide mass range $A \ge 10$.
The RPA calculations for nuclei have systematic uncertainty, which is not well known and is discussed in Section~\ref{Conclusion}.
This uncertainty can be originated from that the nuclear force is not well understood. 
The RPA calculation itself for electron systems reproduces the photoabsorption spectra with satisfactory accuracy~\cite{Nakatsukasa01}.
The experimental data also has systematic uncertainty, as introduced in this section.
We found that there is a systematic difference between the experimental data and the RPA calculations for $E1$ strength, which is shown in Section~\ref{ImpactCross}.
In this study, we demonstrated how the uncertainties affect the resulting UHECR energy spectra and compositions.
The uncertainties will be studied in both theoretical and experimental ways.

In Section~\ref{Method}, we explain how we calculated the $E1$ strength functions with RPA and how we simulated propagation of UHECR nuclei using CRPropa~3~\cite{CRPropa}.
In Section~\ref{ImpactCross}, the photoabsorption cross sections obtained by the RPA calculations are compared with the cross sections obtained using previous experimental data.
The difference of simulated UHECR energy spectra and compositions when photonuclear reactions in CRPropa~3 are replaced with the results of the RPA calculations is shown in Section~\ref{Spectra}. 
In the later sections, we often describe simulations using CRPropa~3 with the default settings as ``CRPropa default" to be distinguished from the RPA calculations. We describe simulations with the RPA calculations as ``Skyrme-RPA" in this paper.
We also compared astrophysical parameters obtained by fitting data from the Auger Collaboration with CRPropa default and Skyrme-RPA. 
The contributions and influences of individual nuclei are discussed in Section~\ref{Spectra}.
Finally, the obtained results are summarized, and the systematic uncertainty is discussed in Section~\ref{Conclusion}.



\section{Method}
\label{Method}

We applied the RPA to obtain the $E1$ strength function in the following way.
The RPA equation is derived as the small-amplitude limit of the time-dependent density functional theory~\cite{DFTtext1,DFTtext2}. 
Under a weak, time-dependent external field $V_\text{ext}(t)$, the transition density $\delta\rho(t)$, which describes the density fluctuation from the ground-state density $\rho_0$, follows the equation 
\begin{equation}
    i \frac{d}{dt}\delta\rho(t) = \left[ h_0, \delta\rho(t)\right] + \left[ V_\text{ext}(t)+ \delta h(t), \rho_0 \right] \,,
\label{TDDFT}
\end{equation}
where $h_0 = h[\rho_0]$ is the single-particle Hamiltonian. The residual field $\delta h(t)$ is induced by density fluctuation, $h[\rho_0+\delta\rho(t)]=h_0+\delta h(t)$. Assuming that $\delta\rho(t)$, $V_\text{ext}(t)$, and $\delta h(t)$ oscillate with a frequency $\omega$ as $\delta\rho(t)=\delta\rho(\omega)e^{-i\omega t}+\delta\rho^\dag(\omega)e^{i\omega t}$, Eq.~(\ref{TDDFT}) is recast to
\begin{equation}
    \omega\,\delta\rho(\omega) = \left[ h_0, \delta\rho(\omega) \right] + \left[ V_\text{ext}(t)+\delta h(\omega), \rho_0 \right] \,.
\label{TDDFT2}
\end{equation}
The residual field $\delta h(\omega)$ is calculated by the functional derivative of the single-particle Hamiltonian with respect to the density, $\delta h = \partial h/\partial \rho \cdot \delta \rho$.
Because $\delta\rho(\omega)$ is not necessarily Hermitian, we introduce forward and backward amplitudes, $|X_i(\omega)\rangle$ and $|Y_i(\omega)\rangle$, to express the transition density $\delta\rho(\omega)$, 
\begin{equation}
    \delta\rho(\omega) = \sum^A_{i=1} \left\{ | X_i(\omega)\rangle\langle \phi_i| + | \phi_i\rangle\langle Y_i(\omega) | \right\} \,,
\end{equation}
where $A=N+Z$ is the mass number of a nucleus, and $|\phi_i\rangle$ are the occupied orbitals in the ground state, $h_0|\phi_i\rangle=\epsilon_i |\phi_i\rangle$ $(i=1,2,\cdots,A)$. Substituting this into Eq.~(\ref{TDDFT2}), we obtain the RPA equations:
\begin{align}
    \omega| X_i(\omega)\rangle &= \left( h_0 - \epsilon_i \right) |X_i(\omega)\rangle + \left\{ V_\text{ext}(t)+\delta h(\omega) \right\} |\phi_i\rangle \,, 
   \label{RPA_X}\\
    -\omega \langle Y_i(\omega)| &= \langle Y_i(\omega)| \left( h_0 - \epsilon_i\right) + \langle \phi_i| \left\{ V_\text{ext}(t) + \delta h^\dag(\omega) \right\}  \,.
  \label{RPA_Y}
\end{align}
In our implementation, we employ the grid representation of the three-dimensional Cartesian-coordinate space.

To calculate the continuous strength function for a given one-body operator $F$, we adopt an external field of $V_\text{ext}(t)=Fe^{-i\omega t}+ F^\dag e^{i\omega t}$. Then, the discretized transition strength is expressed with the forward and backward amplitudes,
\begin{eqnarray}
&&  S(E;F) \equiv \sum_n \left| \langle n | F|0\rangle\right|^2 \delta(E-E_n) \nonumber\\
&&   \phantom{S(E;F)} = - \frac{1}{\pi} \mathrm{Im}\sum_i \left\{ \langle \phi_i |F|X_i(\omega)\rangle + \langle Y_i(\omega) |F^\dag |\phi_i \rangle \right\}     
\end{eqnarray}
for a real frequency $\omega= E$. Here, $|n\rangle$ are energy eigenstates of the total system. 
Since the transition strength is continuous above the nucleon decay energy, we introduce complex frequencies with a finite imaginary part, $\omega= E+i \gamma/2$. 
$\gamma$ describes the spreading width of the GDR and we adopt $\gamma=2.0$~MeV.
The transition strength becomes
\begin{equation}
    S(E;F) = \frac{\gamma}{2\pi} \sum_n \left\{ \frac{\left| \langle n | F|0\rangle\right|^2}{\left( E-E_n\right)^2 + \left( \gamma/2\right)^2} -  \frac{\left| \langle n | F^\dag|0\rangle\right|^2}{\left( E+E_n\right)^2 + \left( \gamma/2\right)^2}  \right\}
    \label{E1cont}
\end{equation}
For the dipole case, we consider an electric dipole operator for $F$:
\begin{equation}
    F=D_z = \frac{N}{A}e \sum^Z_{p=1} z_p - \frac{Z}{A}e\sum^N_{n=1} z_n \,,
\end{equation}
and similar operators for $D_x$ and $D_y$. The photoabsorption cross section is given as
\begin{equation}
    \sigma(E) = \frac{4\pi^2E}{3c}\sum_{\mu=x,y,z}S(E;D_\mu) \,.
\end{equation}

We used three parameter sets of the Skyrme energy-density functional, SkM$^\ast$~\cite{SkMs}, SLy4~\cite{SLy4}, and UNEDF1~\cite{UNEDF1}. SkM$^\ast$ is one of the commonly used Skyrme functionals for nuclear structure calculations. SLy4 is constructed to reproduce the equation of state of infinite nuclear matter proposed by~\cite{APR} as well as experimental data of binding energies and radii in a wide mass region, especially in neutron-rich nuclei. UNEDF1 is recently designed as a sophisticated version of the Skyrme parameter set. For nuclei with odd neutron and/or proton numbers, we employ the filling approximation \cite{Beiner}.
Table~\ref{SepEn} lists the one-nucleon separation energies applied in the calculations.

\begin{figure}
      \centering
      \includegraphics[width=\textwidth]{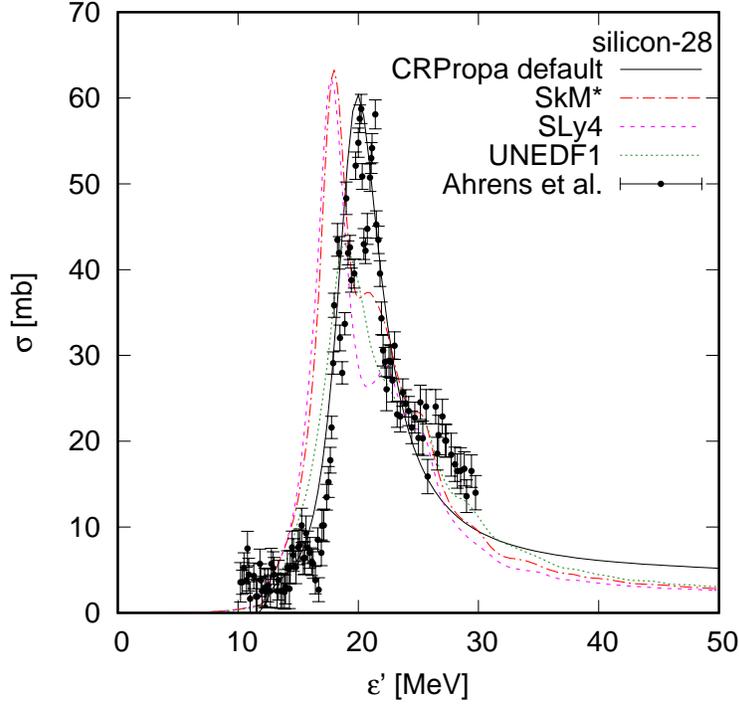}
      \caption{Comparison of nonelastic cross sections of $^{28}$Si of TALYS using $E1$ strength functions of experimental data and models. $\epsilon^{\prime}$ denotes the photon energy in the rest frame of the nucleus. The cross sections at $\epsilon^{\prime} = 0.2, 0.4, 0.6, \ldots 50$ MeV were discretely calculated and plotted with lines. 
      The black solid line was taken from the CRPropa3-data repository~\cite{CRPropa3_data_Repos} and was calculated using the $E1$ strength function with the parameters of the experimental data in Table~2 of CRPropa~3~\cite{CRPropa}. 
      The red dash-dotted, pink long dashed, and forest green dashed lines were derived from the $E1$ strength functions which were calculated using the RPA calculations with different density functionals. The black data points were taken from Ref.~\cite{Ahrens1975} for comparison with models. The data points show measured photoabsorption cross sections of natural isotropic samples of Si.} 
      \label{28Siphotoabs} 
\end{figure}

\begin{figure}
      \centering
      \includegraphics[width=\textwidth]{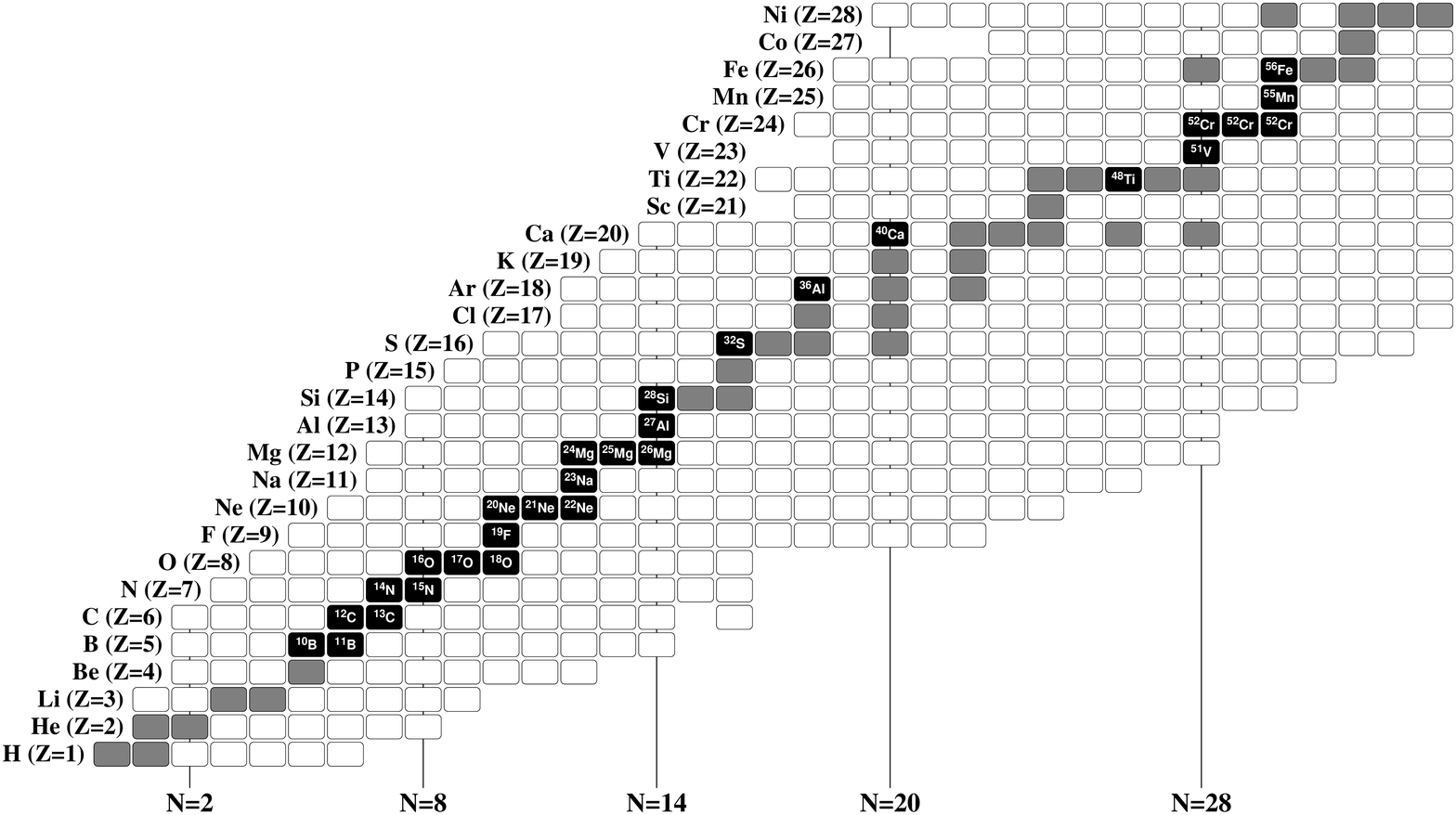}
      \caption{A chart of nuclear species. The horizontal axis represents the number of neutrons, and the vertical axis represents the number of protons. Gray or black boxes show stable nuclei. The 29 black boxes show nuclei that we considered in this work.}
      \label{chart} 
\end{figure}

We used the $E1$ strength function up to 200 MeV obtained from the RPA calculations as an input to version 1.96 of the TALYS code.
Other parameters of TALYS such as the quasi-deuteron component and giant quadrupole resonances are set to the default values.
We obtained nonelastic cross sections as outputs of TALYS\@.
CRPropa~3 reflects the experimental data by using the GDR parameters of the IAEA atlas~\cite{IAEA_Atlas}, if available, as an input to TALYS-1.8.
Fig.~\ref{28Siphotoabs} shows the nonelastic cross sections of $^{28}$Si calculated using different functionals.
The difference of the peak energies and cross sections in this figure mainly affect propagation of UHECR nuclei.
We repeated this procedure for 29 stable nuclei ($^{10}$B, $^{11}$B, $^{12}$C, $^{13}$C, $^{14}$N, $^{15}$N, $^{16}$O, $^{17}$O, $^{18}$O, $^{19}$F, $^{20}$Ne, $^{21}$Ne, $^{22}$Ne, $^{23}$Na, $^{24}$Mg, $^{25}$Mg, $^{26}$Mg, $^{27}$Al, $^{28}$Si, $^{32}$S, $^{36}$Ar, $^{40}$Ca, $^{48}$Ti, $^{51}$V, $^{52}$Cr, $^{53}$Cr, $^{54}$Cr, $^{55}$Mn and $^{56}$Fe). 
Fig.~\ref{chart} shows a chart of nuclear species where the 29 nuclei are highlighted.
Many of these 29 nuclei are expected to be in the decay chain of primary nuclei such as $^{14}$N, $^{28}$Si, and $^{56}$Fe, which are often assumed in the propagation of UHECR nuclei in intergalactic space.

\begin{table}[tbh]
\caption{Separation energies of nuclei in MeV}
\label{SepEn}
\begin{tabular}{cccc}
\hline
Nuclear Species & SkM$^{\ast}$ & SLy4 & UNEDF1 \\
$^{10}$B & 11.2 & 10.5 & 9.5 \\
$^{11}$B & 13.7 & 13.4 & 11.5 \\
$^{12}$C & 14.1 & 14.1 & 12.3 \\
$^{13}$C & 8.7 & 9.5 & 8.3 \\
$^{14}$N & 8.2 & 9.1 & 8.4 \\
$^{15}$N & 10.3 & 11.5 & 10.7 \\
$^{16}$O & 10.3 & 11.2 & 10.1 \\
$^{17}$O & 7.4 & 6.7 & 6.0 \\
$^{18}$O & 7.7 & 7.2 & 5.9 \\
$^{19}$F & 9.1 & 9.2 & 9.0 \\
$^{20}$Ne & 9.2 & 9.3 & 8.4 \\
$^{21}$Ne & 10.8 & 10.6 & 9.8 \\
$^{22}$Ne & 11.0 & 10.7 & 9.2 \\
$^{23}$Na & 10.0 & 9.9 & 9.6 \\
$^{24}$Mg & 9.5 & 9.4 & 8.6  \\
$^{25}$Mg & 10.6 & 10.1 & 9.5 \\
$^{26}$Mg & 11.3 & 11.0 & 9.9 \\
$^{27}$Al & 10.5 & 10.5 & 10.1 \\
$^{28}$Si & 10.3 & 10.5 & 9.5 \\
$^{32}$S & 7.3 & 7.4 & 6.6 \\
$^{36}$Ar & 6.4 & 7.4 & 6.8 \\
$^{40}$Ca & 7.5 & 8.4 & 7.6 \\
$^{48}$Ti & 9.5 & 9.7 & 9.5 \\
$^{52}$Cr & 9.0 & 9.1 & 8.9 \\
$^{56}$Fe & 8.8 & 8.9 & 8.5 \\
\hline
\end{tabular}
\end{table}

\begin{figure}
      \centering
      \includegraphics[width=\textwidth]
      {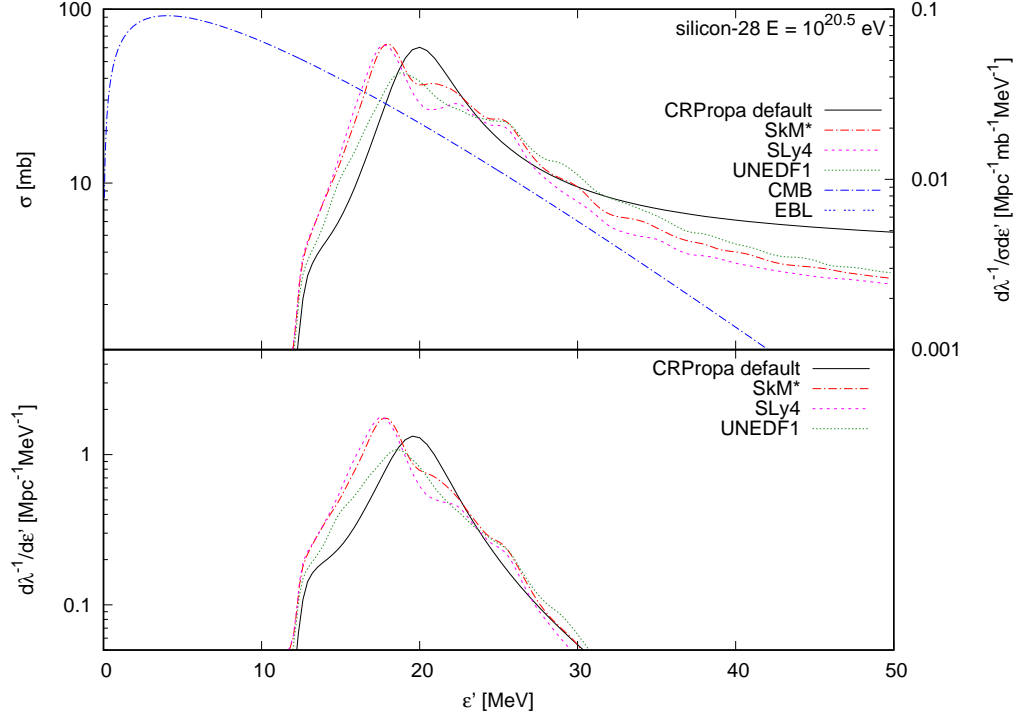}
      \caption{Top panel: the black solid, red dash-dotted, pink long dashed, and forest green dashed lines show the nonelastic cross sections of $^{28}$Si in Fig.~\ref{28Siphotoabs} using logarithmic scales. The left vertical axis shows the cross section. $\epsilon^{\prime}$ shows photon energy in the rest frame of the nucleus. The blue dash-dotted line denotes the differential coefficients of the reaction rates $d\lambda^{-1}/d\epsilon^{\prime}$ of $^{28}$Si with $E = 10^{20.5}$ eV divided by the nonelastic cross sections $\sigma$. The right vertical axis shows the scale of the blue line. The blue line was obtained using the cosmic microwave background (CMB) photons, extragalactic background light (EBL) 
      photons~\cite{Gilmore} and the Lorentz factor of $^{28}$Si. 
      Contribution of the EBL photons is outside of the range in this figure.
      Bottom panel: the differential coefficients of the reaction rates $d\lambda^{-1}/d\epsilon^{\prime}$ of $^{28}$Si with $E = 10^{20.5}$ eV obtained by multiplying nonelastic cross sections $\sigma$ with $d\lambda^{-1}/\sigma d\epsilon^{\prime}$ in the top panel are plotted. The black solid, red dash-dotted, pink long dashed, and forest green dashed lines show different models of the nonelastic cross sections of $^{28}$Si in Fig.~\ref{28Siphotoabs}. The reaction rate $\lambda^{-1}$ of $^{28}$Si with $E = 10^{20.5}$ eV can be obtained by integrating $d\lambda^{-1}/d\epsilon^{\prime}$ over photon energies $\epsilon^{\prime}$.}
      \label{12Csigma} 
\end{figure}

The integral of the nonelastic cross sections with background photons to calculate the reaction rates is given as
\begin{equation}
    \lambda^{-1} (\Gamma, z) = \frac{1}{2 \Gamma^2} \int^{\infty}_{0} \int^{2 \Gamma \epsilon}_{0} n(\epsilon, z) \frac{1}{\epsilon^2} \epsilon^{\prime} \sigma(\epsilon^{\prime}) d\epsilon^{\prime} d\epsilon,
    \label{intrate}
\end{equation}
where $\lambda$ is the mean free path, $\Gamma$ is the Lorentz factor of the nucleus, $z$ is the redshift, $n$ is the spectral number density of background photons, $\epsilon$ is the photon energy, $\epsilon^{\prime} = \Gamma \epsilon (1 - \cos \theta)$ is the photon energy in the rest frame of the nucleus, $\theta$ is the opening angle between the photon and the nucleus momenta, $\sigma$ is the nonelastic cross section.
This formula was shown as Eq.~(4) in~\cite{SteckerSalamon}.
The differential coefficient of Eq.~(\ref{intrate}) can be given by 
\begin{equation}
    \frac{d\lambda^{-1} (\Gamma, z)}{d\epsilon^{\prime}} = \sigma(\epsilon^{\prime})\frac{\epsilon^{\prime}}{2 \Gamma^2} \int^{\infty}_{\frac{\epsilon^{\prime}}{2 \Gamma}} \frac{n(\epsilon, z)}{\epsilon^2} d\epsilon.
    \label{intratediff}
\end{equation}
We used this formula to illustrate the impact of the nonelastic cross sections and background photon energy spectrum on the reaction rates.
Fig.~\ref{12Csigma} shows the integral of the photon spectrum in the right formula of Eq.~(\ref{intratediff}) and the differential coefficient $d\lambda^{-1} (\Gamma, z)/d\epsilon^{\prime}$, respectively.
A summary of the nonelastic cross sections and the mean free paths is shown in Section~\ref{ImpactCross}.

We simulated the energy spectra and compositions of UHECR nuclei on the earth using CRPropa~3 and fitted the results of the Pierre Auger Collaboration using the phenomenological model developed by themselves~\cite{AugerFit}.
The work in Ref.~\cite{AugerFit} was preceded by preliminary results in~\cite{Augerfit_preced}, and the preliminary update was presented in~\cite{Augerfit_update, Augerfit_update2}. 
The following things were assumed in the simulations.
There are identical UHECR sources distributed in the universe, which emit five nuclei, i.e., $^1$H, $^4$He, $^{14}$N, $^{28}$Si, and $^{56}$Fe.
The injection energy spectrum of these nuclei is described by 
\begin{equation}
    \frac{dN}{dE} = C f_A \left( \frac{E}{10^{18} {\rm eV}} \right)^{-p} f_{\rm cut} (E, ZR_{\rm cut}), 
    \label{InjSpec}
\end{equation}
where $C$ is the normalization constant, $f_A$ is the relative fraction of the nucleus, $p$ is the power index of the energy spectrum and $f_{\rm cut} (E, ZR_{\rm cut})$ is given by
\begin{equation}
    f_{\rm cut} (E, ZR_{\rm cut}) = 
    \begin{cases}
        1, & E < ZR_{\rm cut} \\
        \exp(1 - E/ZR_{\rm cut}), & E \ge ZR_{\rm cut}
    \end{cases}
    \label{fcut}
\end{equation}
where $R_{\rm cut}$ denotes the cutoff rigidity.
Eqs.~(\ref{InjSpec}) and (\ref{fcut}) were also taken from Eqs.~(2.1) and (2.2) in Ref.~\cite{AugerFit}.
The evolution of the emissivity of sources is $(1 + z)^{m}$ per unit comoving volume.
The Gilmore 2012 model~\cite{Gilmore} is used for the extragalactic background light.
Magnetic fields are assumed to be small enough to consider one-dimensional cosmic ray propagation from extragalactic sources to the earth.
In the simulations, the step sizes of $p$, $\log (R_{\rm cut})$ and $m$ are set to be 0.1, 0.1 and 1, respectively. The minimum value of $m$ is set to be 0.
The shift of the energy scale of the model $\Delta E/E$ is used when the simulated spectra and compositions at the energy $(E + \Delta E)$ are compared with the data at the energy $E$. $\Delta E/E$ was considered as a nuisance parameter. The distribution of $\Delta E/E$ was assumed to be uniform from -0.14 to +0.14 as in Ref.~\cite{DESY}.

We took $C$, $f_A$ of five nuclei, $p$, $R_{\rm cut}$, $m$, and the energy scale as free parameters.
Only four $f_A$ of five nuclei are independent because the total $f_A$ is fixed to be 1.
We used fifteen data points of the energy spectrum, ten data points of $\langle \ln A \rangle$ and ten data points of $\sigma^2 \left( \ln A \right)$ above $10^{18.7}$ eV for fitting assuming one of the hadron interaction models, SIBYLL-2.3c~\cite{Sibyll} for simplicity.  
All of the data points can be found in the website of the Pierre Auger Observatory~\cite{Auger_public}.
``Combined Spectrum data 2019'' and ``$X_{\rm max}$ and $\ln \left( A \right)$ moments 2019'' are used in this paper.
The energy range $E > 10^{18.7}$ eV is the same as the previous studies~\cite{DESY, AugerFit}.
In fitting the energy spectrum, $\langle \ln A \rangle$ and $\sigma^2 \left( \ln A \right)$, we considered only the error bars of statistical uncertainties as Gaussian errors. 
We searched the energy scale as one of the fitting parameters within 14\% systematic uncertainty of the Pierre Auger Observatory~\cite{Auger}.
In fitting $\langle \ln A \rangle$ and $\sigma^2 \left( \ln A \right)$, we did not consider systematic uncertainties because the systematic uncertainties are not simple to evaluate accurately, and because the fit is used in this work only to show the effect of photodisintegration uncertainties.
For similar reasons, we neglect any possible correlations between $\langle \ln A \rangle$ and $\sigma^2 \left( \ln A \right)$.
The systematic uncertainties of $X_{\rm max}$ are directly related to $\langle \ln A \rangle$ and are asymmetric and slightly energy dependent~\cite{Augerfit_update}.
Consistency of this fitting procedure was checked with Ref.~\cite{DESY}.
In Ref.~\cite{DESY}, the best-fit parameters ($p = -0.8, R_{\rm cut}({\rm V}) = 10^{18.2}, m = 4.2$, and $\Delta E/E =+14\%$) were obtained using {\textsc{PriNCe}} code with SIBYLL-2.3 and TALYS by a combined fit of Auger 2017 data.
We obtained the best-fit parameters ($p = -1.0, R_{\rm cut}({\rm V}) = 10^{18.2}, m = 4$, and $\Delta E/E =-8\%$) using CRPropa3 code with SIBYLL-2.3 and TALYS by fitting Auger 2017 data, and some differences from Ref.~\cite{DESY} in the best-fit parameters would be due to the difference of propagation codes.
Then, we replaced the reaction rates of 29 nuclei with the calculated ones using the RPA calculations in CRPropa~3, simulated the propagation of UHECR nuclei, and repeated the fit.

\section{Model dependence of cross sections of the GDRs} 
\label{ImpactCross}

The GDR is the collective excitation of atomic nuclei, in which all protons and all neutrons oscillate out of phase.
The GDR exhausts almost all of the total cross section of the dipole excitations. 
As it is a fundamental vibration mode of nuclei, GDR properties have been investigated for a long time and have served 
as an alternative observable correlated with other properties of nuclei. For example, the dipole polarizability, which can be evaluated from the cross section of dipole modes, has been applied for constraining the nuclear equation of state~\cite{Polar1, Polar2, Polar3}.

The approximate total cross section is calculated analytically, known as the Thomas-Reich-Kuhn sum rule value.
However, the cross section distribution depends on the density functional, as shown in Fig.~\ref{28Siphotoabs}.
This is because these density functionals have slightly different spin-isospin and momentum dependence.
For instance, the energies of the peak cross section of $^{28}$Si listed in Table~\ref{28Si_peak_energy} have discrepancies of approximately 1 MeV, and they also deviate from the TALYS prediction.
The discrepancy of the calculated peak energies and experimental one has been a long-standing problem that must be overcome, but it is difficult because there are not clear relations of the model parameters to the peak energy.
Hence, it is important to see how these differences impact UHECRs.

\begin{table}[tbh]
\caption{Peak energy of calculated cross section of $^{28}$Si in MeV}
\label{28Si_peak_energy}
\begin{tabular}{ccc}
\hline
 SkM$^{\ast}$ & SLy4 & UNEDF1 \\
18.0 & 17.8 & 18.9 \\
\hline
\end{tabular}
\end{table}

\begin{figure}
      \centering
      \includegraphics[width=\textwidth]{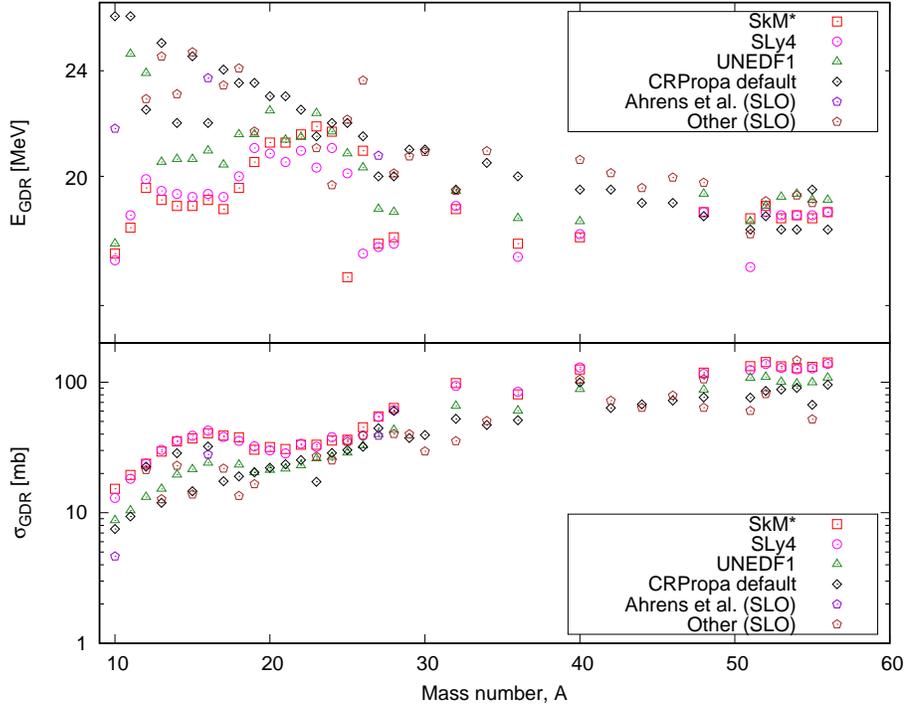}
      \caption{Top panel: peak energies of the giant dipole resonances. Only the first peaks of the resonances with higher cross sections are plotted. The black diamond shows the cross section~\cite{CRPropa3_data_Repos} used in CRPropa~3. The violet and brown hexagon show the standard Lorentzian model (SLO) of the experimental data in the IAEA library~\cite{IAEA}. The photoabsorption cross sections were directly measured using accelerator facilities in Mainz for the red hexagons, and indirectly measured for the yellow hexagons. $^{14}$C, $^{40}$Ar and $^{48}$Ca were not plotted to avoid mixing up with $^{14}$N, $^{40}$Ca and $^{48}$Ti, respectively. Red square, pink circle, and forest green triangle show the cross sections obtained using the different density functionals in the RPA calculations. Bottom panel: photoabsorption cross sections of the giant dipole resonances at the peak energies.}
      \label{Peak} 
\end{figure}

\begin{figure}
      \centering
      \includegraphics[width=\textwidth]{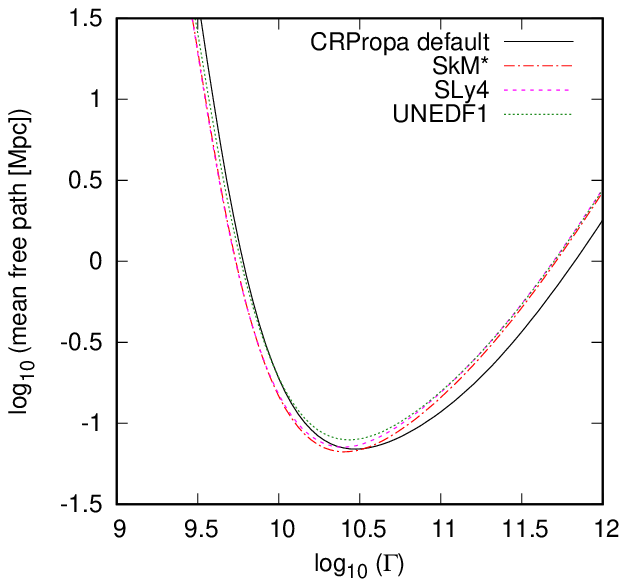}
      \caption{Model predictions of mean free paths using the cross sections of Fig.~\ref{28Siphotoabs} with cosmic microwave background photons when redshift $z = 0$. $\Gamma$ is a Lorentz factor of a $^{28}$Si nucleus.}
      \label{28Sirate} 
\end{figure}

The top panel and bottom panel of Fig.~\ref{Peak} show the peak energies and cross sections of the GDRs, respectively.
There are systematic differences between models; the RPA calculations with SkM$^{\ast}$ and SLy4 tend to underestimate the peak energies and overestimate the peak cross sections in light nuclei. 
The mean free paths directly reflect the trend.
Fig.~\ref{28Sirate} shows mean free paths of $^{28}$Si obtained using the nonelastic cross sections in Fig.~\ref{28Siphotoabs}.
The difference of the peak energies of the GDRs in Fig.~\ref{28Siphotoabs} affects the Lorentz factors where the GDRs are the main channels in Fig.~\ref{28Sirate}, and the difference of the peak cross sections in Fig.~\ref{28Siphotoabs} affects the mean free paths in Fig.~\ref{28Sirate}. 

\section{Model dependence of simulations of propagation of UHECRs}
\label{Spectra}

We fitted the experimental data of the Pierre Auger Observatory using CRPropa default and Skyrme-RPA with the method shown in Section~\ref{Method} and obtained the fit parameters in Table~\ref{FitParams}.
The explanations of the parameters in Table~\ref{FitParams} are the followings. 

The parameters $p$, $R_{\rm cut}$, $m$, $\Delta E/E$, and $f_A$ were defined in Section~\ref{Method}.
$\chi^{2}_{\rm spec.}$ shows $\chi^2$ values of fitting energy spectrum, and $\chi^{2}_{\rm comp.}$ shows the summation of $\chi^2$ values of fitting $\langle \ln A \rangle$ and $\sigma^2 \left( \ln A \right)$.
Fifteen data points were used for fitting the energy spectrum, and twenty data points were used for fitting the $\langle \ln A \rangle$ and $\sigma^2 \left( \ln A \right)$.
Eight free parameters were used, and the degrees of freedom of the combined fit is twenty-seven.
$\chi^{2}_{\rm spec.} + \chi^{2}_{\rm comp.}$ values in Table~\ref{FitParams} are much higher than the degrees of freedom.
The fit is used in this work only to exemplify the effect of photodisintegration uncertainties, so possible reasons for the poor goodness of fit and ways to improve it are outside the scope of this work.
In this paper, we want to focus on the difference caused by the photodisintegration models that occur at the highest energies, so the higher $\chi^2$ values at lower energies do not impact the conclusions of this paper.
Many of the fit parameters in Table~\ref{FitParams} show significant differences between CRPropa default and Skyrme-RPA. 
The GDR peaks of Skyrme-RPA appear at lower energies than CRPropa default, and the difference in the GDR peaks mainly causes relatively soft UHECR spectra with Skyrme-RPA, as shown later in this section. 
In particular, $f_A (^{28} \rm Si)$ and $f_A (^{56} \rm Fe)$ of Skyrme-RPA are larger than CRPropa default to compensate the difference of the energy spectra in Table~\ref{FitParams}.
The proton fraction $f_A (^1 \rm H)$ cannot be determined because of the limited contribution above $10^{18.7}$ eV, so the proton fraction in Table~\ref{FitParams} is fixed to be 0.
The step sizes of $p$, $\log (R_{\rm cut})$ and $m$ are comparable to the uncertainties or smaller than those, so the errors of these parameters are not described in Table~\ref{FitParams}.

\begin{figure}
      \centering
        \includegraphics[width=\textwidth]{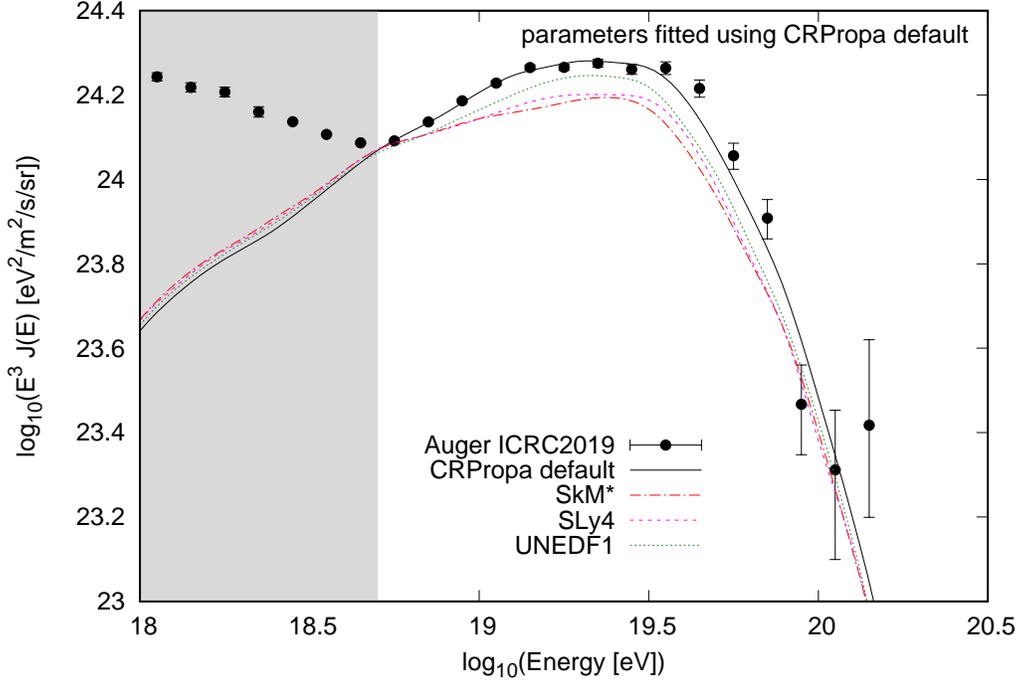}
      \caption{Comparison of simulated energy spectra and observed energy spectrum by the Pierre Auger Collaboration~\cite{Auger}. 15 data points of the energy spectrum above $10^{18.7}$ eV, 10 data points of $\langle \ln A \rangle$ and 10 data points of $\sigma^2(\ln A)$ were fitted with simulated results. We simulated energy spectra with different models using best-fit parameters of CRPropa default, listed in the column of CRPropa default in Table~\ref{FitParams}.}
      \label{CompEspectraNoNorm} 
\end{figure}

\begin{figure}
      \centering
      \includegraphics[width=\textwidth]{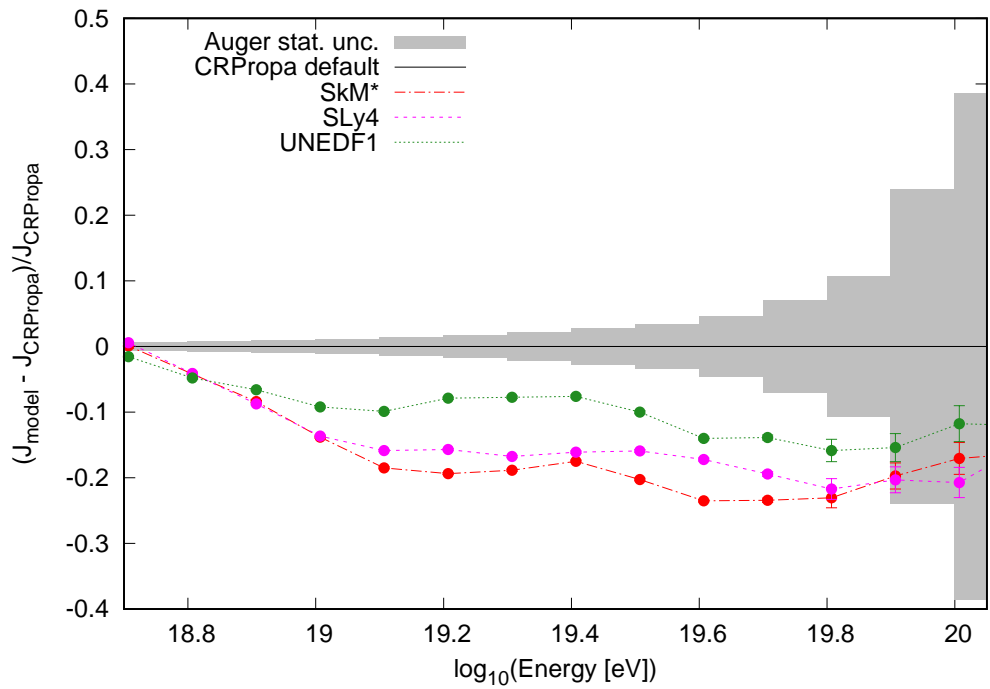}
      \caption{The relative differences of the energy spectra with the RPA calculations $J_{\rm{model}}$ from CRPropa default $J_{\rm{CRPropa}}$ are plotted. $J_{\rm{model}}$ and $J_{\rm{CRPropa}}$ are the same as Fig.~\ref{CompEspectraNoNorm}. The statistical uncertainties of the data are shown as the hatched region.}
      \label{CompEspectraNoNormError} 
\end{figure}

\begin{figure}
      \centering
      \includegraphics[width=\textwidth]{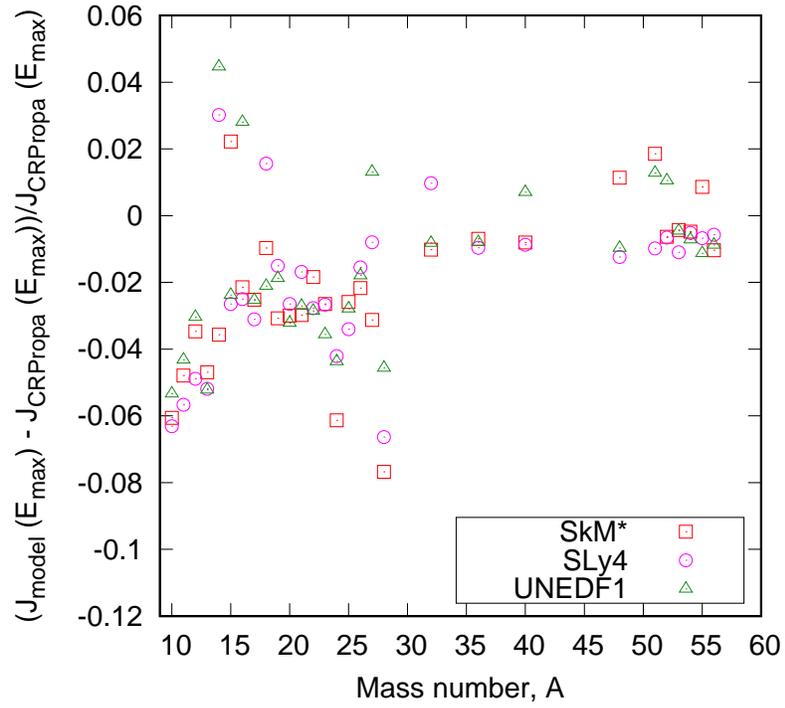}
      \caption{Each point was computed using CRPropa default for all nuclei except one with mass number $A$ computed using the RPA calculations. The vertical axis shows maximum differences from CRPropa default in the spectral shape. $E_{\rm max}$ denotes the energy that shows the maximum difference in the spectral shape. $J_{\rm{model}}$ and $J_{\rm{CRPropa}}$ represent the same spectra as Fig.~\ref{CompEspectraNoNormError}.}
      \label{MaxDiffEspec} 
\end{figure}

\begin{figure}
      \centering
      \includegraphics[width=\textwidth]{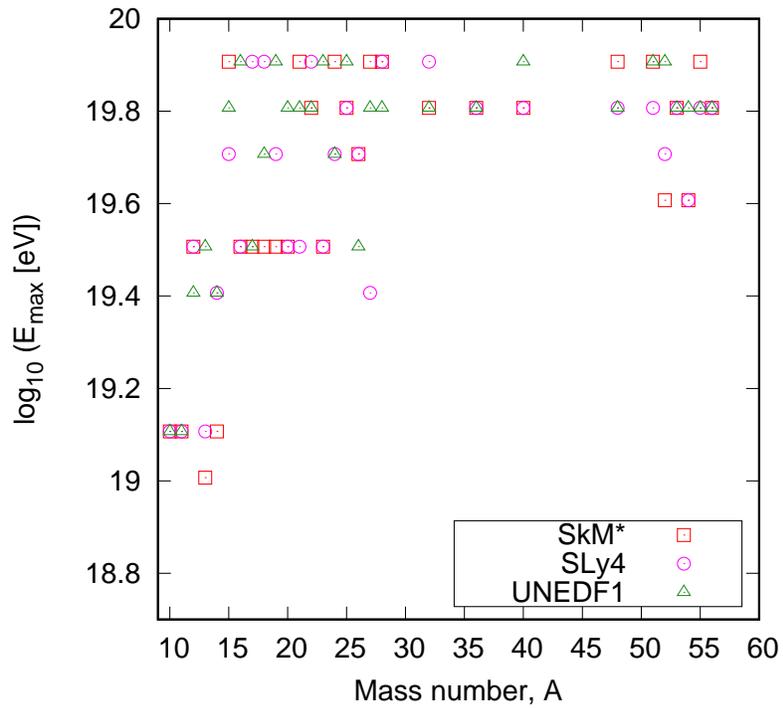}
      \caption{The vertical axis denotes energies that show the maximum difference in the spectral shape in Fig.~\ref{MaxDiffEspec}. The step size of searched energies is 0.1 in $\log (E/$eV). $E > 10^{19.9}$ eV is not searched because of the limited number of simulated events.}
      \label{MaxDiffEspecPeakEn} 
\end{figure}

\begin{figure}
      \centering
      \includegraphics[width=\textwidth]{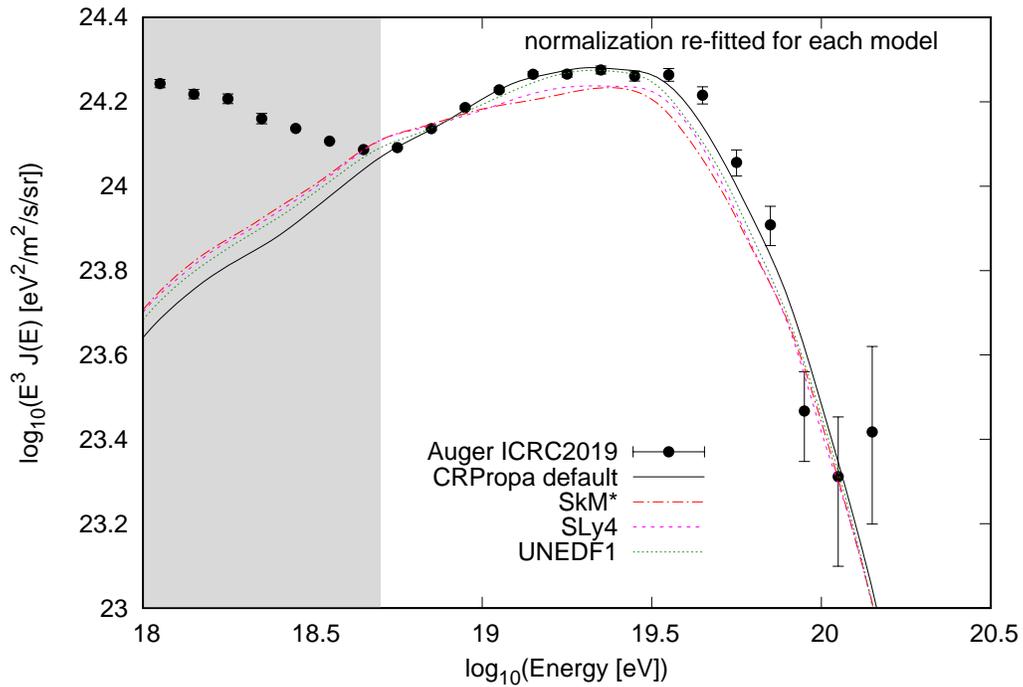}
      \caption{The same figure as Fig.~\ref{CompEspectraNoNorm} except for the normalizations of the red, pink, and forest green lines, obtained by fitting the data points with $E > 10^{18.7}$ eV. The red solid, pink long dashed, and forest green dashed line are inconsistent with the data, and $\chi^2$ values with these lines and the data points above $10^{18.7}$ eV are 472, 366 and 116, respectively.}
      \label{CompEspectra} 
\end{figure}

\begin{figure}
      \centering
      \includegraphics[width=\textwidth]{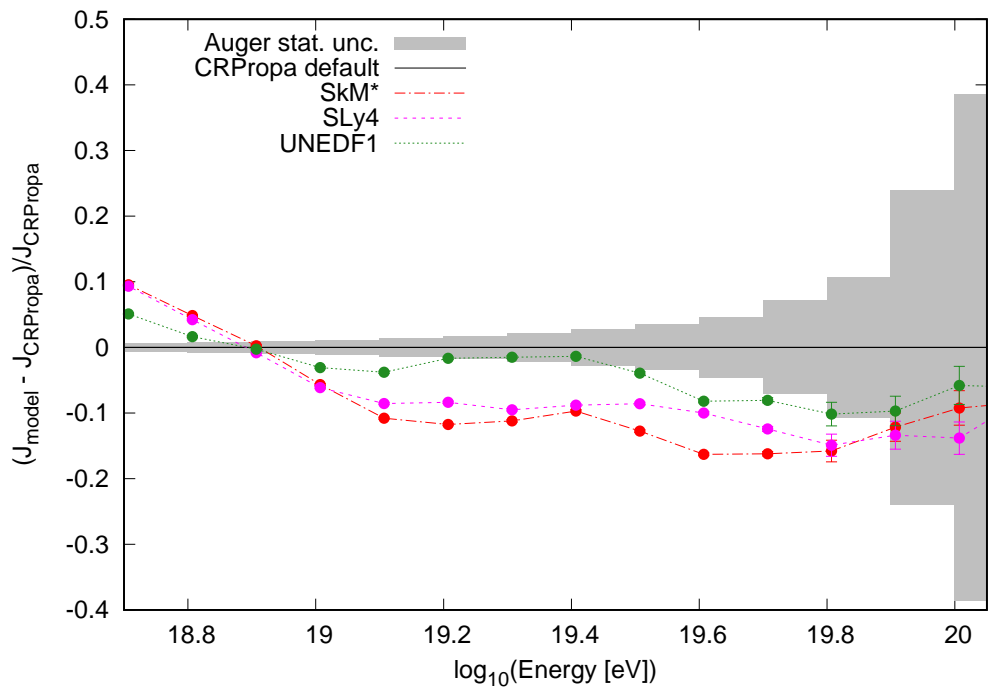}
      \caption{The same results as Fig.~\ref{CompEspectra} are plotted here. In this figure, the relative differences of the model predictions of the intensities are plotted as Fig.~\ref{CompEspectraNoNormError}.}
      \label{CompEspectraError} 
\end{figure}

\begin{figure}
      \centering
      \includegraphics[width=\textwidth]{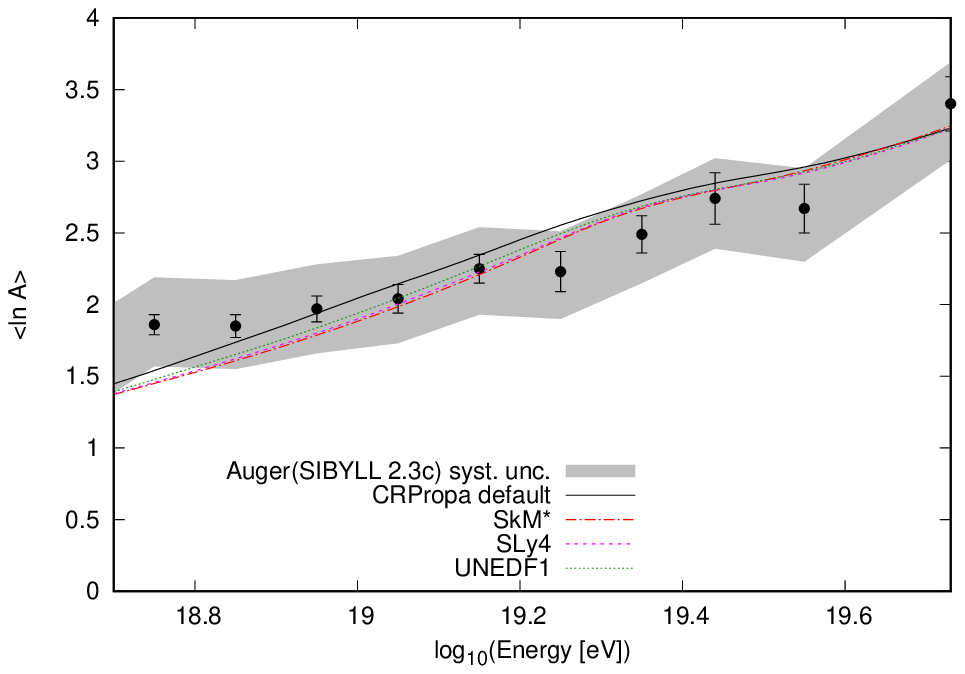}
      \caption{Comparison of simulated $\langle \ln A \rangle$ of the same results as Fig.~\ref{CompEspectra}.}
      \label{lnA} 
\end{figure}

\begin{figure}
      \centering
      \includegraphics[width=\textwidth]{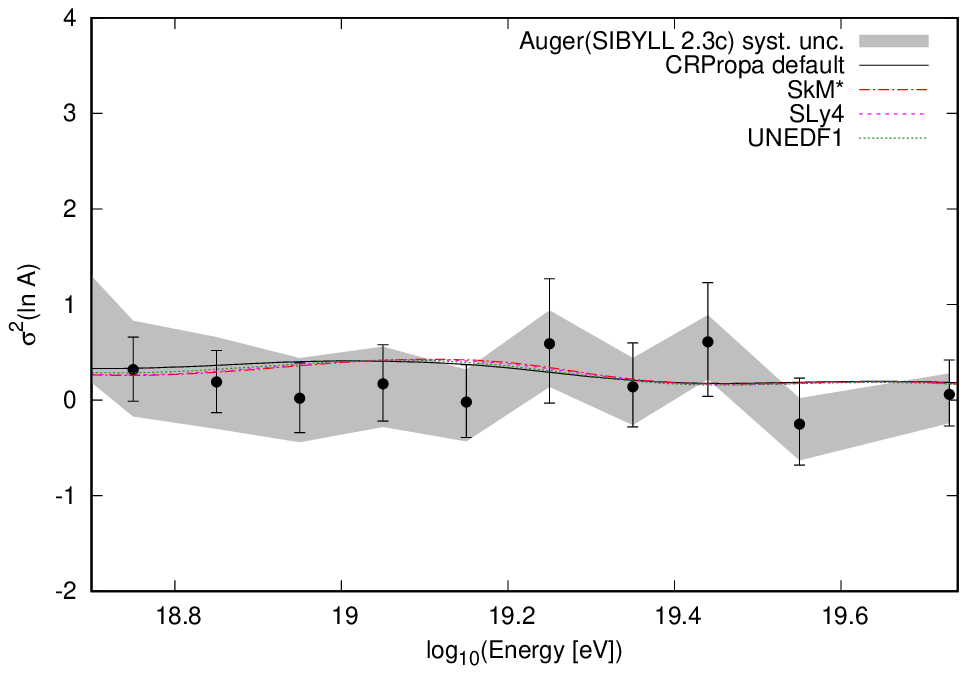}
      \caption{Comparison of simulated $\sigma^{2} (\ln{ A})$ of the same results as Fig.~\ref{CompEspectra}.}
      \label{SigmalnA} 
\end{figure}

\begin{figure}
      \centering
      \includegraphics[width=\textwidth]{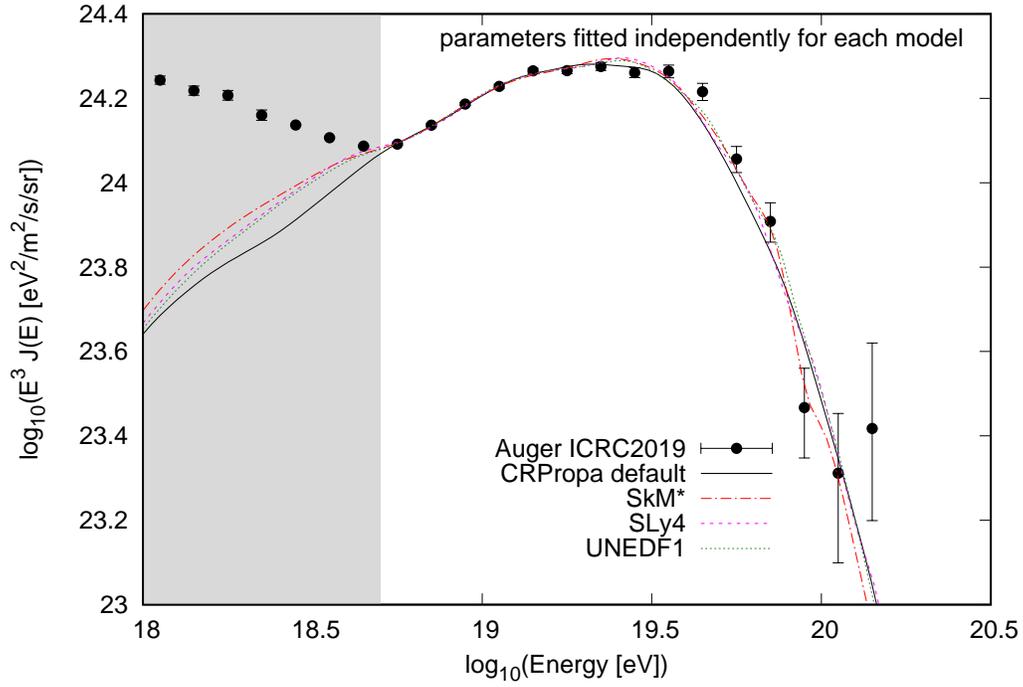}
      \caption{The data points and simulated energy spectrum with CRPropa default are the same as Fig.~\ref{CompEspectra}. 
      The red dash-dotted line denotes CRPropa default in this figure. In Fig.~\ref{CompEspectraNoNorm} and Fig.~\ref{CompEspectra}, the black solid line was used for CRPropa default to emphasize the model used for the fit.
      The other plotted energy spectra were simulated using Skyrme-RPA with their best-fit astrophysical parameters. The parameters are listed in Table~\ref{FitParams}. The parameters were obtained by fitting the observed energy spectrum, $\langle \ln A \rangle$ and $\sigma^{2} (\ln{ A})$. }
      \label{CompEspectra_paramfit} 
\end{figure}

\begin{figure}
      \centering
      \includegraphics[width=\textwidth]{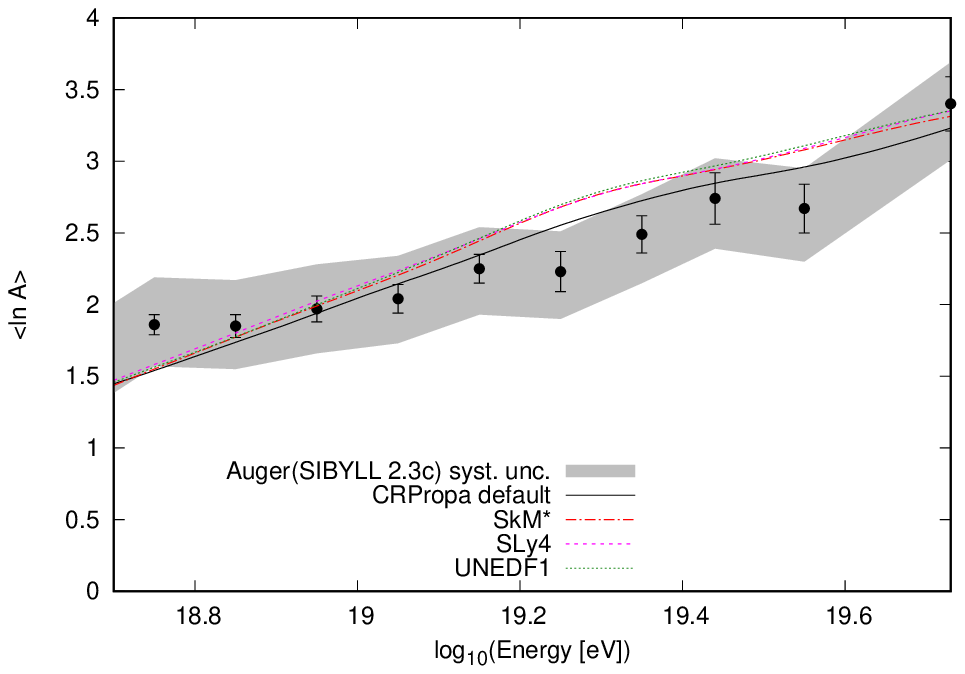}
      \caption{Comparison of simulated $\langle \ln A \rangle$ of the same results as Fig.~\ref{CompEspectra_paramfit}.}
      \label{ComplnA_paramfit} 
\end{figure}

\begin{figure}
      \centering
      \includegraphics[width=\textwidth]{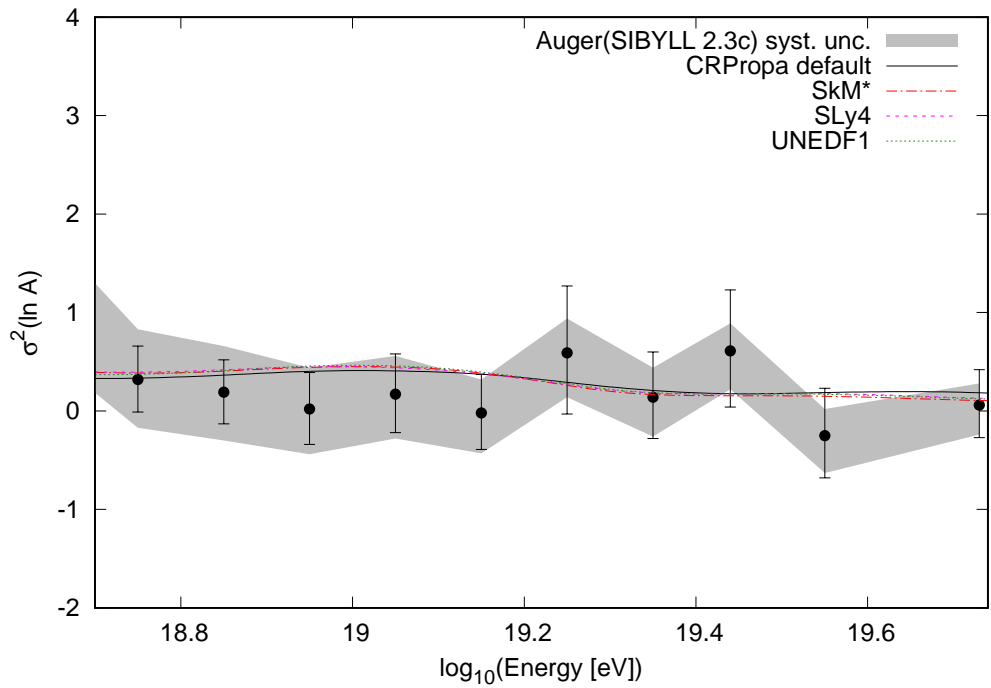}
      \caption{Comparison of simulated $\sigma^{2} (\ln{ A})$ of the same results as Fig.~\ref{CompEspectra_paramfit}.}
      \label{CompSigmalnA_paramfit} 
\end{figure}

Fig.~\ref{CompEspectraNoNorm} and~\ref{CompEspectraNoNormError} show the comparison results of the energy spectra, which were calculated with CRPropa default and Skyrme-RPA using the same parameters in the column of CRPropa default in Table~\ref{FitParams}. We did not change the total number of simulated events between CRPropa default and Skyrme-RPA in these figures.
The maximum difference of the energy spectra between CRPropa default and Skyrme-RPA is more than 20\%.
The difference is more than the statistical uncertainty of the experimental data, as shown in Fig.~\ref{CompEspectraNoNormError}.
We repeated the comparison of energy spectra when the RPA calculations of only one nucleus are applied to understand the contribution of which nucleus is important.
Then, we searched the energy bins to find the maximum difference between CRPropa default and Skyrme-RPA in the spectral shape.
Fig.~\ref{MaxDiffEspec} shows the maximum differences from default CRPropa in the spectral shape when the RPA calculations of only one nucleus are applied.
The largest maximum difference from CRPropa default is obtained when the RPA calculations of $^{28}$Si are applied. 
The difference can be more than 5\%, depending on the density functional.
Fig.~\ref{MaxDiffEspecPeakEn} shows the energy bins, showing the maximum difference in the spectral shape. 
The lighter nuclear species show the maximum difference at the lower energies because the Lorentz factors in the GDRs are relatively larger for lighter nuclei. 
We also fitted the normalization of the energy spectra with $E > 10^{18.7}$ eV which were calculated with the RPA calculations in order to estimate the difference's impact on the interpretation in Fig.~\ref{CompEspectra} and~\ref{CompEspectraError}. 
The maximum difference of the energy spectra between CRPropa default and Skyrme-RPA can be more than 20\% which is more than the statistical uncertainty of the experimental data even if the normalization is adjusted.
We compared the model predictions of $\langle \ln A \rangle$ and $\sigma^2 (\ln A)$ in Fig.~\ref{lnA} and Fig.~\ref{SigmalnA}, respectively.
The differences of $\langle \ln A \rangle$ and $\sigma^2 (\ln A)$ between the models are smaller than the systematic uncertainty of the experimental data.
Fig.~\ref{CompEspectra_paramfit}, Fig.~\ref{ComplnA_paramfit}, and Fig.~\ref{CompSigmalnA_paramfit} show comparison between the data and models with their best-fit parameters in Table~\ref{FitParams}. 

Different photonuclear reaction models that lead to different fit parameters were already noticed in the previous works~\cite{DESY, AugerFit, Augerfit_update}. 
Section 5.2 in Ref.~\cite{DESY} showed the difference in the fit parameters between TALYS, PSB, and PEANUT models. Section 5.3.3 in Ref.~\cite{AugerFit} showed the difference between TALYS, PSB, and Geant4~\cite{Geant4} models, and Section 4 in Ref.~\cite{Augerfit_update} showed the difference between TALYS and PSB models. 
As shown in Section 5.3.3 in Ref.~\cite{AugerFit}, the PSB cross sections neglect $\alpha$-particle productions and imply larger cutoff rigidity than TALYS. The difference in the cross sections also affected elemental fractions obtained by a combined fit between the PSB model and TALYS. 
The photonuclear reaction models in the previous works are different from the RPA calculations shown in this work.
We found that the RPA calculations imply lower cutoff rigidity than the CRPropa default that uses TALYS (see Fig.~\ref{CompEspectraNoNorm} and Fig.~\ref{CompEspectraNoNormError}), mainly because of the difference in the GDR peaks.
Therefore, the RPA calculations tend to have the opposite effect to the PSB model compared with TALYS, reflecting the difference in the cutoff rigidity.

\begin{table}[tbh]
\caption{Fit parameters}
\label{FitParams}
\begin{tabular}{ccc}
\hline
 Fit parameter & CRPropa default & SkM$^\ast$ \\
 $p$ & -1.0 & -0.9  \\
 $R_{\rm cut}$ (V) & $10^{18.1}$ & $10^{18.2}$ \\
 $m$ (evolution parameter) & 1 & 0 \\
 Shift of the energy scale $\Delta E/E$ (\%) & 14$^{+0}_{-3}$ & -10$^{+2}_{-3}$ \\
 $f_A$ ($^1$H) (\%) & 0 & 0 \\
 $f_A$ ($^4$He) (\%) & 94.8$\pm 0.4$ & 93.9$^{+0.7}_{-0.6}$ \\
 $f_A$ ($^{14}$N) (\%) & 5.0$\pm 0.4$ & 5.6$\pm 0.6$ \\
 $f_A$ ($^{28}$Si) (\%) & (1.9$\pm 0.3$) $\cdot$ 10$^{-1}$ & (4.9$\pm 0.7$) $\cdot$ 10$^{-1}$ \\
 $f_A$ ($^{56}$Fe) (\%) & (4.8$^{+1.4}_{-1.1}$) $\cdot$ 10$^{-3}$ & (8.9$^{+3.0}_{-2.3}$) $\cdot$ 10$^{-3}$ \\ \hline
 $\chi_{\rm spec.}^2$ & 27 & 27 \\
 $\chi_{\rm comp.}^2$ & 42 & 65 \\
 ($\chi_{\rm spec.}^2 + \chi_{\rm comp.}^2$)/$n_\text{dof}$ & 69/27 & 92/27 \\
\hline
\\
\hline
 Fit parameter & SLy4 & UNEDF1 \\
 $p$ & -0.7 & -0.7 \\
 $R_{\rm cut}$ (V) & $10^{18.2}$ & $10^{18.2}$ \\
 $m$ (evolution parameter) & 0 & 0 \\
 Shift of the energy scale $\Delta E/E$ (\%) & -6$\pm 2$ & -6 $\pm 2$ \\
 $f_A$ ($^1$H) (\%) & 0 & 0 \\
 $f_A$ ($^4$He) (\%) & 92.9$\pm 0.7$ & 95.2$\pm 0.4$ \\
 $f_A$ ($^{14}$N) (\%) & 6.6$\pm 0.7$ & 4.4$\pm 0.4$ \\
 $f_A$ ($^{28}$Si) (\%) & (5.3$^{+0.9}_{-0.8}$) $\cdot$ 10$^{-1}$ & (4.2$\pm 0.6$) $\cdot$ 10$^{-1}$ \\
 $f_A$ ($^{56}$Fe) (\%) & (1.7$\pm 0.4$) $\cdot$ 10$^{-2}$  & (1.3$\pm 0.3$) $\cdot$ 10$^{-2}$ \\ \hline
 $\chi_{\rm spec.}^2$ & 37 & 23 \\
 $\chi_{\rm comp.}^2$ & 61 & 68 \\
  ($\chi_{\rm spec.}^2 + \chi_{\rm comp.}^2$)/$n_\text{dof}$ & 98/27 & 91/27 \\
\hline

\end{tabular}
\end{table} 


\section{Conclusions and discussions}
\label{Conclusion}

The PANDORA project finally plans to systematically measure photonuclear reactions from light to $A \sim 60$ nuclei.
Nuclear theories are needed to model the experimental data and estimate reactions that are not measured.
The PANDORA project will start to measure the cross sections of a few nuclei in a few years. 
A reliable nuclear theory would also be informative in interpreting the current experimental data before various nuclei are newly measured.
We studied the impact of the difference between CRPropa default and Skyrme-RPA on the interpretation of observations of UHECR nuclei.
TALYS in CRPropa reflects the current experimental data.
We found that the difference between CRPropa default and Skyrme-RPA of 29 stable nuclei in the spectral shape is more than 20\% and larger than the statistical uncertainty of the observed UHECR energy spectrum when the same astrophysical parameters are assumed.
The systematic model dependence of the GDR peaks was the biggest contribution to the difference.
The biggest contribution is provided by $^{28}$Si in the spectral shape.
The difference can be more than 5\% if the model of $^{28}$Si is changed.
The difference of best-fit astrophysical parameters obtained by a combined fit of UHECR energy spectrum and composition data can also be more than the uncertainty of the data between CRPropa default and Skyrme-RPA, as shown in Table~\ref{FitParams}.
The uncertainty of UHECR simulations can also propagate to the dispersion of simulated secondary neutrino and gamma-ray fluxes as demonstrated in~\cite{Batista19}.
In the current situation, systematic uncertainties can exist in both theory and experiment, and it is difficult to conclude the source of the difference.
The following things are considered to improve the experimental data and RPA calculations.

There were systematic discrepancies in the cross sections between experimental facilities in the past. The PANDORA project plans to solve such systematic uncertainty using modern experimental methods at three independent facilities.

To improve the systematic model dependence of the GDR peaks, the density functional of the RPA calculations needs to be corrected because the density functional is not currently tuned to reproduce properties of excited states.
Improving the Skyrme density functional model parameters is not straightforward because there is no one-to-one correspondence between each parameter and the GDR's peak energy and height. On the other hand, quite recently, the Monte Carlo calculation shows that some of the Skyrme parameters are correlated with the peak energy of the GDR~\cite{inakura22}. The correlation would enable us to improve the Skyrme parameters. Therefore, we are planning to propose a new Skyrme parameter set that reproduces the experimental data on $E1$ strength and gives more reliable $E1$ strength of nuclei in which there are no experimental data in the near future.

\section*{Acknowledgement}
This work was supported by JSPS KAKENHI (A) Grant Number JP19H00693. 
We are thankful to members of the PANDORA project for fruitful discussions. The simulations in this work were partially obtained using HOKUSAI supercomputer at RIKEN. E.K. and S.N. thank supports from "Pioneering Program of RIKEN for Evolution of Matter in the Universe (r-EMU)". We are also thankful to Donald Warren III for many corrections of the sentences.


\bibliography{mybibfile}

\end{document}